# Graduate teaching assistants use different criteria when grading introductory physics vs. quantum mechanics problems

Charles Henderson,[1] Emily Marshman,[2] Ryan Sayer,[3] Chandralekha Singh,[2] Edit Yerushalmi[4]

[1] *Department of Physics, Western Michigan University, 1903 W. Michigan Ave., Kalamazoo, MI, 49008*
[2] *Department of Physics and Astronomy, University of Pittsburgh, 3941 O'Hara St., Pittsburgh, PA 15260*
[3] *Department of Physics, Bemidji State University, Bemidji, MN, 56601*
[4] *Department of Science Teaching, Weizmann Institute of Science, 234 Herzl St., Rehovot, Israel 7610001*

**Abstract.** Physics graduate teaching assistants (TAs) are often responsible for grading. Physics education research suggests that grading practices that place the burden of proof for explicating the problem solving process on students can help them develop problem solving skills and learn physics. However, TAs may not have developed effective grading practices and may grade student solutions in introductory and advanced courses differently. In the context of a TA professional development course, we asked TAs to grade student solutions to introductory physics and quantum mechanics problems and explain why their grading approaches were different or similar in the two contexts. TAs expected and rewarded reasoning more frequently in the QM context. Our findings suggest that these differences may at least partly be due to the TAs not realizing that grading can serve as a formative assessment tool and also not thinking about the difficulty of an introductory physics problem from an introductory physics student's perspective.

## I. INTRODUCTION

Physics graduate Teaching Assistants (TAs) often grade student work in introductory and advanced courses at large universities. The TAs' beliefs about grading and grading practices can shape student learning and communicate instructors' goals and expectations to students [1]. Goals for many physics courses at all levels include helping students learn physics and develop effective problem solving skills. Findings of physics education research (PER) suggest that placing the burden of proof for explicating the problem solving process on students in both intro and advanced courses can promote these goals [2]. Since grading can play an important role in learning, the TAs can benefit from opportunities to reflect upon their beliefs about grading and their grading practices in both intro and advanced courses.

However, most TAs receive very little training or guidance regarding grading [3]. Their grading beliefs and practices are often based upon their own experiences as students [4]. TAs may not think of grading as a tool for formative assessment, which can promote learning. It is also possible that many TAs perceive the difficulty of the problem they are grading from their own perspective instead of the perspective of their students and they use different criteria for grading intro and advanced students' work [5]. For example, the TAs have significantly more expertise in solving intro physics problems, and they may not think about the difficulty of an intro physics problem from their students' perspective. They may assume that the intro physics answers are obvious and so intro students do not need to show their work while solving problems [5]. As a result, when grading intro physics solutions, the TAs may ignore solution features that are conducive to learning for intro students and may not require that students explicate the problem solving process. On the other hand, since TAs may not yet be experts in advanced courses like quantum mechanics (QM), they may perceive a QM problem to be difficult. As a result, it is possible that when grading QM student solutions, the TAs can identify solution features that are necessary for conveying understanding in the QM context and expect students to explicate the problem solving process. Therefore, they may be stricter in grading QM than intro solutions.

We investigated these hypotheses and whether physics graduate TAs grade student solutions in intro physics and QM using different criteria and the reasons for possible differences. The findings can serve as a useful tool for leaders of professional development courses who are interested in helping TAs improve their grading practices.

## II. METHODOLOGY AND FINDINGS

This case study involved 15 first-year physics graduate TAs participating in their first semester in a mandatory, semester-long professional development course at a research university in the U.S. There were 11 male and 4 female TAs. The majority of them were teaching recitations for intro physics courses for the first time. A few were also assigned to facilitate a laboratory section or grade students' work in various physics courses. The professional development course met for 2 hours each week and was meant to prepare the TAs for their teaching responsibilities. The TAs were generally asked to do one hour of homework each week pertaining to teaching that was graded for completeness. During class meetings, TAs usually discussed their homework assignment from the previous week in small groups. At the end of the class, they shared what they had discussed in groups while the instructor provided feedback.

One sequence of homework and in-class activities pertained to grading. At the beginning of the semester, TAs were given an intro physics problem (see Fig. 1) and Student Solution D (SSD) and Student Solution E (SSE) (see Fig. 2) to that problem. They were given a research-validated worksheet that asked them to grade the student solutions SSD and SSE out of 10 points in a quiz context, list solution

You are whirling a stone tied to the end of a string around in a vertical circle having a radius of 65 cm. You wish to whirl the stone fast enough so that when it is released at the point where the stone is moving directly upward it will rise to a maximum of 23 meters above the lowest point in the circle. In order to do this, what force will you have to exert on the string when the stone passes through its lowest point one-quarter turn before its release? Assume that by the time you have gotten the stone going and it makes its final turn around in the circle, you are holding the end of the string at a fixed position. Assume also that air resistance can be neglected. The stone weighs 18 N. The correct answer is 1292 N.

**FIGURE 1**. The intro physics problem for which the TAs graded student solutions SSD and SSE.

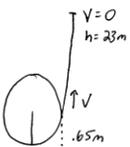

**FIGURE 2**. For the intro physics problem, Student Solution D (SSD) and Student Solution E (SSE).

features, and justify the weight they assigned to each solution feature to arrive at a final score [6,7]. A week after the TAs worked through the grading activities involving the intro physics student solutions, they were given a QM problem along with Student Solution 1 (SS1) and Student Solution 2 (SS2) (see Fig. 3) to the QM problem. The TAs were provided with a correct instructor's solution to the QM

For an electron in a one-dimensional infinite square well with well boundaries at $x = 0$ and $x = a$, measurement of position yields the value $x = a/2$. Write down the wave function immediately after the position measurement and without normalizing it show that if energy measured immediately after the position measurement, it is equally probable to find the electron in any odd-energy stationary state.

**FIGURE 3**. The upper-level QM problem for which the TAs graded solutions SS1 and SS2.

**FIGURE 4**. For the QM problem, Student Solution 1 (SS1) and Student Solution 2 (SS2).

problem and a worksheet that asked them to grade the student solutions to the QM problem out of 10 points in a quiz context, list solution features, and justify the weight they assigned to each solution feature to arrive at a final score (similar to the grading activity for intro physics). The TAs were also asked whether they used different criteria when grading student solutions in the QM context vs. the intro physics context and if they did so, why was that the case.

The grading worksheets were previously developed and validated by three of the authors in collaboration with two graduate student researchers in physics education for use with TAs/instructors [6,7]. The intro physics student solutions SSD and SSE [6] and QM student solutions were created based upon common student responses to the problems and were iterated many times between researchers and other physics instructors. All of the student solutions have the correct final answer. However, the elaborated intro solution SSD and the QM solution SS2 both explicate the problem solving process. On the other hand, the brief intro solution SSE and QM solution SS1 do not explicate the problem solving process. The contrasting solution features were designed to encourage graders to reflect on various problem solving approaches that educational literature suggests promote desired problem-solving practices [2,7,8].

After an initial analysis of the collected data, seven of the TAs in the study volunteered to be interviewed to provide further clarification of their grading beliefs and practices and to investigate whether the grading activities carried out in the TA training class impacted their beliefs about their grading in some manner not captured in their written responses. The interview protocol included a set of pre-determined questions. The interviewer also asked follow-up questions on-the-spot to examine the TAs' reasoning about interview responses as well as responses on the worksheets.

The TAs' assigned scores on the QM solution and the intro physics solution were analyzed. Table I shows the average scores and standard deviations when TAs graded the intro solutions and the QM solutions in the quiz context. TAs tended to grade the elaborated solutions higher than the brief solutions in both the intro and QM contexts, but the difference was more pronounced for the QM solutions than for the intro physics solutions. The highest disagreement among the TAs was about scores to assign to the brief solution of the intro problem SSE (the standard deviation was 2.71). We performed $t$-tests for comparison, and found that the differences in averages were statistically significant between the QM solutions SS1 and SS2 ($p = 0.008$) but not statistically different for the two intro solutions ($p = 0.274$).

To investigate the reasons for why the TAs graded on different criteria in the two contexts, we analyzed the TAs' stated reasons for why they graded differently in the two contexts. Data sources were written responses, class discussions, and interviews. The TAs were asked to write responses to the following two questions in the worksheet involving the QM grading activity: 1. "Was your grading approach different when grading intro physics student solutions vs. upper-level QM student solutions? Why or why not?" 2. "How did your grading considerations change when grading intro physics student solutions vs. upper-level QM student solutions? What are the reasons for the differences?"

The TAs' written responses about the reasons for why they would grade differently (or not do so) in the two contexts were analyzed using open-coding to generate initial categories grounded in the actual data. Once initial categories were agreed upon, the coding was completed by two of the researchers separately. After comparing codes, any disagreements were discussed and the categories were refined until better than 90% agreement was reached. Table II shows the categories of the TAs' written responses for why (or why not) they would grade differently in the intro physics and QM contexts and the percentages of the TAs who mentioned each category. We note that the TAs could have written more than one reason for why they graded differently in both the QM and intro physics solution contexts.

**TABLE I**. Average (Avg.) quiz scores assigned to the brief and elaborated (Elab.) solutions to the intro and QM physics problems, with standard deviations (S.D.)

|  | Intro Physics | | QM | |
| --- | --- | --- | --- | --- |
|  | Brief (SSE) | Elab. (SSD) | Brief (SS1) | Elab. (SS2) |
| Avg. | 7.07 | 7.93 | 6.57 | 8.47 |
| S.D. | 2.71 | 1.24 | 2.06 | 1.55 |

In their written responses, 10 of the 15 TAs (67%) noted that they would grade the QM and intro physics problems differently, while 5 TAs (33%) noted that they would not grade differently in the two contexts. (However, an analysis of their actual grading shows that 3 out of 5 of these TAs also graded the two contexts differently.) Below, we discuss TAs' stated reasons for why they graded the QM and intro physics solutions using different criteria and graded the QM solutions more strictly than the intro student solutions.

Out of the 15 TAs in the course, twenty-six percent of them expected that students should explicitly demonstrate their understanding when solving QM problems as opposed to intro physics problems because QM problems are more complex. For example, one TA gave this explanation for why he would use different criteria: "For the upper level courses, the concepts are more complex, need more explanation." Interviews suggest that the TAs with these types of responses were able to gauge the difficulty of a QM problem from the perspective of an advanced student because they were themselves at a similar expertise level. In interviews, some of the TAs explicitly mentioned that QM problems are significantly more difficult than intro physics problems. One interviewed TA explained that she had thought about how much more difficult a QM problem is, and therefore focused significantly more on the proof of understanding when grading solutions to QM problems, stating: "In QM I don't expect people to be able to do things in their mind, so if they're not writing it down I kind of feel they don't know it." The TAs who felt that the QM problem is significantly more difficult than the intro physics problem often stated that they expected the students in QM courses to explicate the problem solving process and they would grade them on the use of a systematic approach to problem solving that includes a conceptual analysis of the physics problem (but they did not have the same high expectations when grading intro solutions). Some interviewed TAs who claimed that QM is more difficult than intro physics mentioned that intro physics is easier than QM because intro physics is more concrete. For example, one TA mentioned "For intro physics we can make an example to understand the questions more clearly." Another TA compared herself to her students (instead of putting herself in intro students' shoes and thinking from intro students' perspective) and stated: "In intro physics, since I can do it in my mind, I think that intro students can do it [in their minds] too." Responses of this type indicate that some TAs may not have thought about the difficulty of the intro problem from the perspective of an intro student [5] and may not expect intro students to explicate the problem solving process when solving intro problems (and do not penalize them for not doing it). However, intro students may find the intro physics problem challenging because it is sufficiently abstract for them similar to advanced students struggling with a QM problem. Similar to previous studies [9,10], we find that the TAs, who thought that the intro problems are easy, often inferred correct understanding in student's solution SSE when there was no evidence of it.

**Table II.** Categories of TAs' stated reasons for grading QM solutions differently than intro physics solutions and the percentages (%) of TAs in each category ($N = 15$ TAs).

| Category | % |
| --- | --- |
| Demonstrating understanding is more important in QM than intro physics because the subject is more complex | 26 |
| Demonstrating understanding is more important in QM than intro physics because it is expected of advanced students | 26 |
| Grading should focus more on conceptual understanding in QM and more on use of equations, calculations, solving steps, correct math and final answer in intro physics | 40 |

In their written responses, 26% of the TAs claimed that advanced students should demonstrate their understanding when solving QM problems because they are already expected to have learned physics concepts as well as effective problem-solving approaches. For example, one interviewed TA stated that "high-level students have gone through many years of training, what they need is [to] interpret the problem [to get credit]." This TA felt that after many years of training, students should be able to articulate their thought processes explicitly in their solution in order to receive credit. In addition, 40% of the TAs noted that they would grade advanced students on the explication of concepts, but that they would grade intro students on whether they used correct formulas or mathematical steps and got the correct answer. For example, one interviewed TA stated, "If a student is majoring in physics, they should be able to understand all the concepts perfectly to be able to solve complicated problems. In upper-level courses, I think the student should understand everything they are doing, they are not allowed to just use an equation because they have seen people use [that equation] before." This TA emphasized that a formula-fitting approach was acceptable in intro physics courses but not in advanced physics courses for physics majors. Another TA claimed that her grading focused more on concepts in QM and that "in intro physics (assuming the students are not majoring in physics) it's okay if they only learn how to use equations and how to solve problems because they might have not seen physics problems before in their life…." A TA who only valued clarification and interpretation in problem solution in the context of QM stated, "I grade more on the interpretation of problems in QM. For intro, I consider more the calculation…"

TAs' belief that they should be stricter in grading in QM than in intro physics because intro students had not learned problem solving skills and were not experts in physics (so a lenient grading standard should be appropriate for them) suggests that they had not thought about grading practices serving as a formative assessment tool. Many of the TAs emphasized grading intro students only on the final answer, correct formulas and mathematical steps because they felt that intro students were not experts in physics. Interviews and discussions suggest that TAs generally viewed grading as only serving as a summative assessment tool in a course at any level and they thought that its only purpose was to determine how much students knew at a given point of time [1]. Since written responses, class discussions and interviews suggest that the TAs had not reflected on how good grading practices can serve as a formative assessment tool [1], they need support in understanding that grading practices can help students develop problem solving skills and learn physics.

### III. SUMMARY

In this study we found that graduate TAs expected and rewarded reasoning in student problem solutions more frequently in a QM context than in an intro physics context. This finding may at least partly be due to the TAs not realizing that grading can serve as a formative assessment tool and not thinking about the difficulty of a problem from their students' perspective. The grading activity involving intro physics and QM solutions can be a valuable tool for investigating TAs' grading beliefs and practices. Leaders of professional development courses can ask TAs to reflect on any differences in their grading of the intro and QM solutions along with the PER recommendation (about the importance of grading on explication of the problem solving process). This reflection may help TAs transition towards desirable grading practices in all courses.